\documentclass[aps,prc,groupedaddress,showpacs,manuscript]{revtex4}

\usepackage{graphicx}

\begin{document}

\title{Probing the momentum dependence of medium modifications of the nucleon-nucleon elastic cross sections}

\author {Qingfeng Li,$\, ^{1}$\footnote{Corresponding author (e-mail: liqf@hutc.zj.cn)}, Caiwan Shen,$\, ^{1}$ and M. Di Toro,$\, ^{2}$}
\address{
1) School of Science, Huzhou Teachers College, Huzhou 313000, People's Republic of China\\
2) Laboratori Nazionali del Sud INFN, I-95123 Catania, Italy }


\begin{abstract}

The momentum dependence of the medium modifications on
nucleon-nucleon elastic cross sections is discussed with microscopic
transport theories and numerically investigated with an updated
UrQMD microscopic transport model. The semi-peripheral Au+Au
reaction at beam energy $E_b=400A$ MeV is adopted as an example. It
is found that the uncertainties of the momentum dependence on medium
modifications of cross sections influence the yields of free
nucleons and their collective flows as functions of their transverse
momentum and rapidity. Among these observables, the elliptic flow is
sensitively dependent on detailed forms of the momentum dependence
and more attention should be paid. The elliptic flow is hardly
influenced by the probable splitting effect of the neutron-neutron
and proton-proton cross sections so that one might pin down the mass
splitting effect of the mean-field level at high beam energies and
high nuclear densities by exploring the elliptic flow of nucleons or
light clusters.

\hspace{2cm}

\end{abstract}

 \maketitle

It is well-known that the extensive explore of the equation of state
(EOS) of the nuclear matter is one of the hottest topics during a
long period of time. With continuous improvements in the reduction
of the stiffness uncertainty of the EOS endeavored in recent years
\cite{Li:2008gp,Fuchs:2007vt,Danielewicz:2002pu,Liu:2001iz}, much
more attention have been kept on going to the more consistent
treatments of the mean field and the collision term, which both
originate from the same effective Lagrangian density based on the
QHD theory \cite{Li:2003vd,Mao:1994zza,Mao:1997gr}.

In the past, two-body nucleon-nucleon (NN) cross sections adopted in
the microscopic transport models are often treated to be in free
space partly for simplicity and partly for the lack of the
information of medium modifications of cross sections. Recently,
various medium modifications on the NN elastic and inelastic cross
sections have been investigated with various theories and simulated
with transport models by several groups. And vivid effects of these
modifications on dynamics of the heavy ion collisions (HICs) have
been found in quite a few sensitive observables
\cite{Li:2003vd,Sammarruca:2005ch,Sammarruca:2005tk,Li:2005jy,Larionov:2003av,Prassa:2007zw}.
Based on QHD-II type effective Lagrangian, in which the interaction
between nucleons is described by exchanges of $\sigma$, $\omega$,
$\pi$, $\rho$ \cite{Li:2000sha} and $\delta$
\cite{Li:2003vd,Li:2003ai} mesons, the in-medium neutron-proton,
proton-proton and neutron-neutron elastic scattering cross sections
($\sigma^*_{np}$, $\sigma^*_{pp}$, and $\sigma^*_{nn}$) had been
systematically studied within the framework of the self-consistent
RBUU transport theory \cite{Mao:1997gr,Mao2005}. Further, in Ref.\
\cite{Li:2006ez}, such medium modifications of the NN elastic cross
section on several observables was investigated within neutron-rich
intermediate-energy HICs. At that time, the so-called splitting
effect of the effective neutron and proton masses (`NR' and `Dirac'
modes) was paid more attention and the HICs at lower beam energies
($\sim 100A$ MeV) were in use. We found that, although the
transverse flow as a function of rapidity and the nuclear stopping
quantities, such as $Q_{zz}$ as a function of momentum and the ratio
of halfwidths of the transverse to that of longitudinal rapidity
distribution $R_{t/l}$, are very sensitive to the medium
modifications of the cross sections, the mass-splitting effect on
these observables is quite small and deserves more investigations.
We notice that some of these findings were also demonstrated
independently by the other theoretical group \cite{Li:2005jy}.

On the mean field level, the sensitive probes to the controversial
mass-splitting effect at intermediate beam energies ($\sim 400A$
MeV) have been studied \cite{DiToro:2006hd,DiToro:2008zm}. It is
found that the elliptic flow, especially the flow difference of free
neutrons and protons (or light isobars), is quite sensitive to the
mass-splitting indicated by various theories. In view of the same
origin of the medium modifications on both the mean field and the
collision, one might ask: might the splitting effect on the two-body
NN (elastic) cross sections also be seen from the elliptic flow
observable? If so, is the total sensitivity to the elliptic flow
reduced or enhanced with the consideration of the splitting effect
in the cross sections? Obviously, the answer to these questions is
essential to determine the trend of the mass-splitting at high
densities.

As well-known from many previous investigations, see, e.g.,
\cite{Mao:1994et,Li:2003vd,Sammarruca:2005ch,Sammarruca:2005tk,Li:2005jy,Li:1993ef,Giansiracusa:1996zz},
the density dependence of cross sections is drastically influenced
by the relative momentum of the colliding particles in the NN
center-of-mass system. In order to consider the momentum dependence
of the medium modifications on cross sections in the real transport
models, some phenomenological scaling models are adopted in previous
calculations \cite{Li:2006ez,Li:2005jy}. However, we know that
different momentum dependent forms of the cross sections can be
obtained based on different parameterizations used in RBUU
calculations \cite{Li:2000sha,Li:2003vd}. In this work, we would
also like to investigate the influence of various momentum
dependence on the observables related to the splitting effect. It is
even more valuable than the splitting effect itself and is necessary
to be resolved in a timely manner since it is comparable to the
density dependence of the cross sections in the nuclear medium and
should influence the determination of the stiffness of the EOS when
comparing same observables with experimental data.

The new-updated UrQMD model, which is suitable for studies of HICs
at SIS energies, is adopted for calculations in this work
\cite{Li:2005gf,Li:2006ez,Li:2008bk}. It is well-known that the
UrQMD microscopic transport model is based analogous principles as
the quantum molecular dynamics model (QMD) \cite{Aichelin:1986wa}
and the relativistic quantum molecular dynamics model (RQMD)
\cite{Sorge:1989dy}: the mean-field potential applied to hadrons is
treated similar to QMD, while the treatment of the collision term is
similar to RQMD. And starting from the version 2.0, the PYTHIA code
is incorporated into UrQMD in order to investigate the jet
production and fragmentation at high SPS and RHIC energies
\cite{Bratkovskaya:2004kv}. Hadrons are represented by Gaussian wave
packets in phase space and the phase space of the hadron is
propagated according to Hamilton¡¯s equation of motion. Besides the
cascade mode, and in terms of better description of the experimental
data, the effective two-body interaction potential terms are taken
into account carefully. In the current version of the UrQMD model
\cite{Li:2005gf,Li:2006ez}, the potential energies include the
two-body and three-body (which can be approximately written in the
form of two-body interaction) Skyrme- (also called as the
density-dependent terms), Yukawa-, Coulomb-, Pauli-,
density-dependent-symmetry-, and momentum-dependent- terms. With the
updates of the UrQMD transport model, some successful theoretical
analyses, predictions and comparisons with data have been
accomplished.

In the previous work \cite{Li:2006ez} the in-medium NN elastic cross
sections $\sigma_{el}^{*}$ are treated to be factorized as the
product of a medium correction factor ($F(u,\alpha,p)$,
$u=\rho/\rho_0$ is the nuclear reduced density and
$\alpha=(\rho_n-\rho_p)/\rho_0$) the isospin-asymmetry) and the free
NN elastic ones $\sigma_{el}^{free}$. For the inelastic channels
$\sigma_{in}$, we still use the experimental free-space cross
sections $\sigma_{in}^{free}$. It is believed that this assumption
does not have serious influence on our present study at intermediate
energies. Therefore, the total two-body scattering cross section of
nucleons $\sigma_{tot}^{*}$ will be modified to
$\sigma_{tot}^{*}=\sigma_{in}+\sigma_{el}^{*}=\sigma_{in}^{free}+F(u,\alpha,p)
\sigma_{el}^{free}$.

\begin{figure}
\includegraphics[angle=0,width=0.6\textwidth]{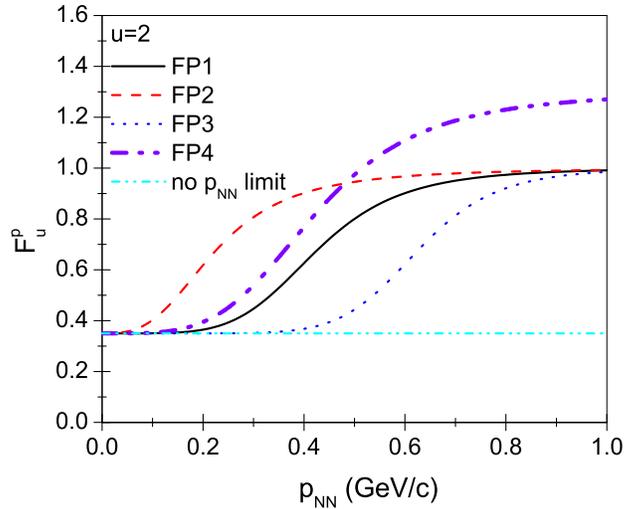}
\caption{ $F_u^p$ as a function of the relative momentum $p_{NN}$
with four parameter sets FP1 $\sim$ FP4 and without $p_{NN}$ limit.
The reduced density $u=2$ is chosen.} \label{fig1}
\end{figure}

As for the medium correction factor $F(u,\alpha,p)$, it is
proportional to both the isospin-scalar density effect $F_u$ and the
isospin-vector mass-splitting effect $F_\alpha$, please read
\cite{Li:2006ez} for more details. Furthermore, the factors $F_u$
and $F_\alpha$ should be constrained by the relative momentum of the
two colliding particles in the NN center-of-mass system ($p_{NN}$).
In \cite{Li:2006ez}, they are formulated as,

\begin{equation}
F_{\alpha,u}^{\rm p}=\left\{
\begin{array}{l}
f_0 \hspace{3.3cm} p_{NN}>1 {\rm GeV}/c \\
\frac{F_{\alpha,u} -f_0}{1+(p_{NN}/p_0)^\kappa}+f_0 \hspace{1cm}
p_{NN} \leq 1 {\rm GeV}/c

\end{array}
\right. . \label{fdpup}
\end{equation}
The parameters $f_0$, $p_0$ and $\kappa$ in Eq.\ (\ref{fdpup}) can
be varied in order to obtain various momentum dependence of, for
example, $F_u$. In this work, we select several parameter sets,
which are shown in Table \ref{tab1}. The corresponding $F_u^p$
functions are illustrated in Fig.\ \ref{fig1} at a reduced density
$u=2$ (and $F_{u=2}=0.35$). The FP1 set was used in our previous
works when the medium modifications of cross sections were
considered. It is also used in this work as a base. The function of
FP2 (FP3) gives a rapid increase at smaller (larger) $p_{NN}$ as
compared to the case with FP1. This demonstrates the uncertainty of
the momentum dependence to the density dependent cross sections.
With a certain set of isospin dependent EOS, the NN elastic cross
section might be even enhanced at large momenta
\cite{Li:2000sha,Zhang:2007gd} (which arises from the differences
between the isoscalar and isovector channels), when compared to the
cross section at free space. FP4 in Fig.\ \ref{fig1} gives one
example to show $30\%$ enhancement at large $p_{NN}$. On the
contrary, the dash-dot-dotted line is to show the case without any
momentum constraint on $F_u$.

\begin{table}
\begin{center}
\renewcommand{\arraystretch}{1.2}
\begin{tabular}{|l|l|c|l|}\hline
\bf Set & $f_0$ & $p_0$ [GeV c$^{-1}$]   & $\kappa$ \\\hline\hline
\tt FP1 & 1   & 0.425 & 5        \\
\tt FP2 & 1   & 0.225 & 3        \\
\tt FP3 & 1   & 0.625 & 8        \\
\tt FP4 & 1.3 & 0.425 & 4        \\
\tt no $p_{NN}$ limit & F(u) & / &/ \\ \hline
\end{tabular}
\end{center}
\caption{\label{stdvector} Four parameter sets FP1 $\sim$ FP4 used
in this work for various momentum dependence of $F_u$. The case
without $p_{NN}$ limit is also considered if one sets $f_0$ to be
F(u) in Eq.\ (\ref{fdpup}).} \label{tab1}
\end{table}

A soft EOS with momentum dependence (SM-EOS) and with a soft
symmetry potential energy (corresponding stiffness factor $\gamma$
of the energy form $S_0 u^\gamma$ is set to $0.5$. $S_0=32$ MeV is
the symmetry energy at the normal density) is adopted in this work.
The reaction Au+Au at a beam energy $E_b = 400A$ MeV and for impact
parameter $b=7$ fm is chosen. For each case 100 thousand events are
calculated and the freeze-out time is taken to be $100 $fm c$^{-1}$.
After freeze-out, a conventional phase-space coalescence model
\cite{Kru85} is used to construct clusters, in which the nucleons
with relative momenta smaller than $P_0$ and relative distances
smaller than $R_0$ are considered to belong to one cluster. In this
work, $P_0$ and $R_0$ are chosen to be $0.3$GeV c$^{-1}$ and $3.5$
fm, respectively. The change of $P_0$ and $R_0$ values certainly
alters the yields of clusters, but it shall not change the main
conclusions drawn in this paper.

\begin{figure}
\includegraphics[angle=0,width=0.8\textwidth]{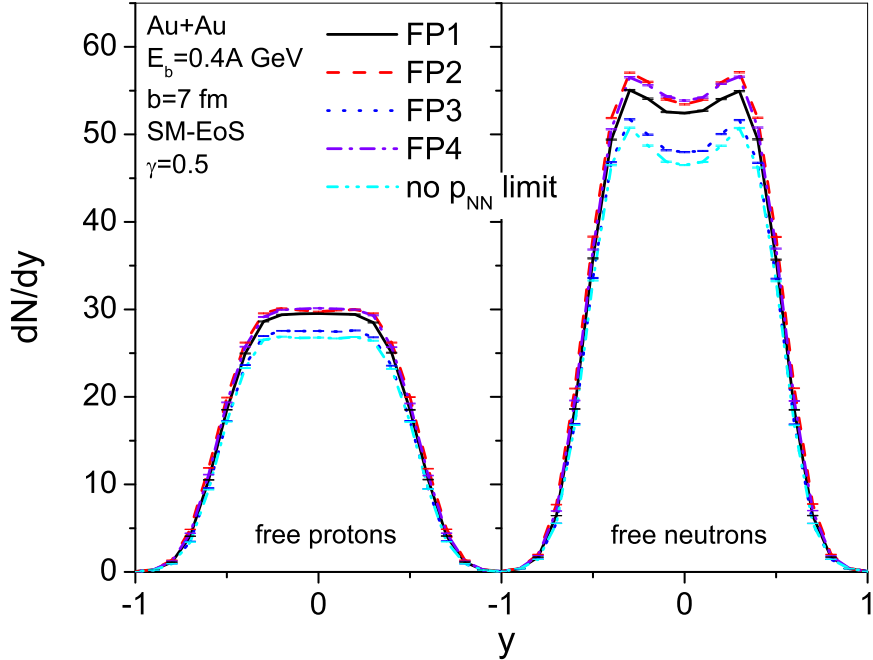}
\caption{Rapidity distribution of protons (left plot) and neutrons
(right plot) for Au+Au reactions at beam energy $E_b = 400A$ MeV and
impact parameter $b = 7$ fm. SM-EOS with soft symmetry potential
energy is adopted for calculations with various (from FP1 to FP4)
and without momentum dependence (``no $p_{NN}$ limit'') of medium
modifications on NN elastic cross sections.} \label{fig2}
\end{figure}

Fig.\ \ref{fig2} shows the rapidity (in the nucleus-nucleus
center-of-mass system) distribution of unbound protons (left plot)
and neutrons (right plot) for Au+Au reactions at beam energy $E_b =
400A$ MeV and impact parameter $b = 7$ fm. It is interesting to see
that in each plot the results can be divided into two camps: The
results with FP1, FP2, and FP4 are similar to each other and
somewhat higher than the results with FP3 and without $p_{NN}$
limit. It implies that at such beam energy the collision dynamics of
nucleons is sensitive to medium modifications of elastic cross
sections in the $0.3\lesssim p_{NN} \lesssim 0.6$GeV c$^{-1}$ region
which is understandable. And, it is also easy to understand that
stronger reduction of the NN elastic cross sections leads to weaker
emission of nucleons. However, in each camp of each plot, it is
still hard to distinguish them by taking only the rapidity
distribution of the yield of nucleons into account and one needs to
go further on.

\begin{figure}
\includegraphics[angle=0,width=0.8\textwidth]{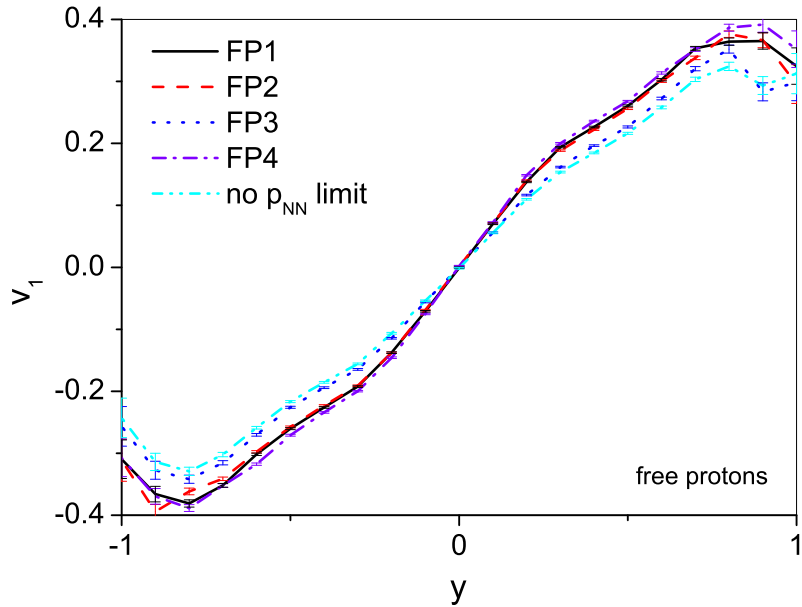}
\caption{Rapidity distribution of the directed flow $v_1$ of protons
for Au+Au reactions at beam energy $E_b = 400A$ MeV and impact
parameter $b = 7$ fm. Results with FP1 $\sim$ FP4 momentum
dependence of medium modifications of cross sections is compared to
the one without momentum constraint.} \label{fig3}
\end{figure}

Fig.\ \ref{fig3} shows the directed flows $v_1$ of protons
($v_1=p_x/p_t$ where $p_t=\sqrt{p_x^2+p_y^2}$ is the transverse
momentum of the particle) as a function of the rapidity. The
uncertainty of momentum dependence on medium modifications of cross
sections is not much obviously seen in the rapidity distribution of
$v_1$ although the two camps shown already in Fig.\ \ref{fig2}
appear again. And, it is clear that larger NN elastic cross sections
in the nuclear medium makes bigger positive directed flow at this
beam energy, which is due to larger transverse expansion.

\begin{figure}
\includegraphics[angle=0,width=0.8\textwidth]{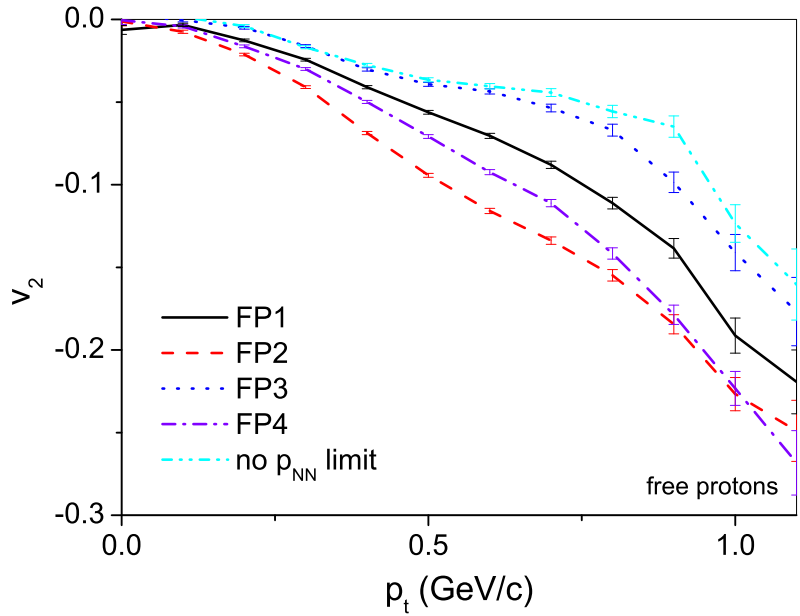}
\caption{Elliptic flows $v_2$ of protons as a function of the
transverse momentum $p_t$.} \label{fig4}
\end{figure}

\begin{figure}
\includegraphics[angle=0,width=0.8\textwidth]{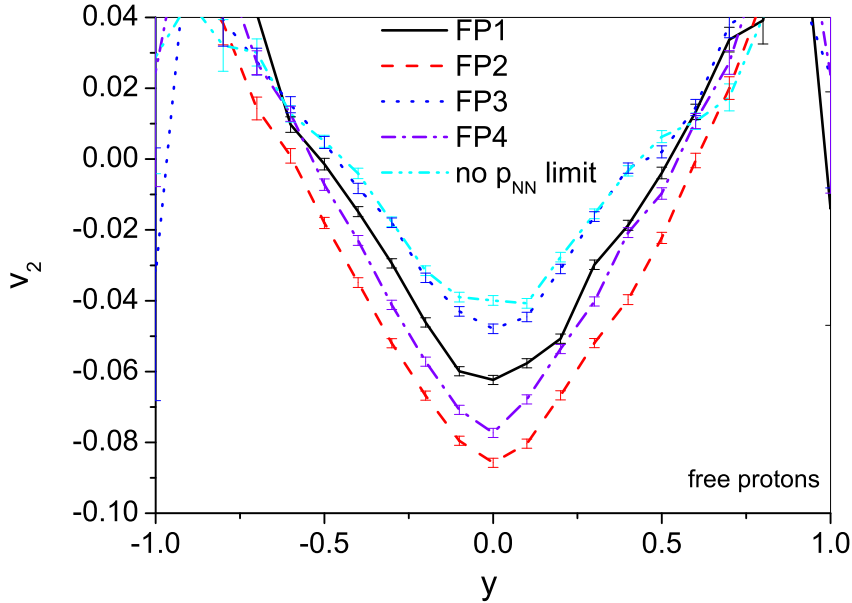}
\caption{Elliptic flows $v_2$ of protons as a function of the
rapidity.} \label{fig5}
\end{figure}

Let us further investigate the sensitivity of the momentum limits to
the elliptic flow $v_2$ ($v_2=<(p_x^2-p_y^2)/p_t^2>$) of protons as
functions of the transverse momentum $p_t$ which is shown in Fig.\
\ref{fig4}, and of the rapidity which is shown in Fig.\ \ref{fig5},
respectively. It is clear that at $0.5 \lesssim p_t \lesssim 1.0$
GeV c$^{-1}$ (in Fig.\ \ref{fig4}) or at mid-rapidity (in Fig.\
\ref{fig5}), the elliptic flow of protons is sensitive to the
treatment of the momentum dependence of medium modifications of
cross sections. It is known that stronger two-body collisions lead
to larger negative elliptic flow at such beam energy, which can be
examined explicitly by relating Fig.\ \ref{fig1} to Figs.\
\ref{fig4} and \ref{fig5}. It is further found that the order of
flows shown in Figs.\ \ref{fig4} and \ref{fig5} follows the momentum
dependent forms in the $0.3 \lesssim p_{NN} \lesssim 0.5$ GeV
c$^{-1}$ region shown in Fig.\ \ref{fig1}. Furthermore, in the $p_t$
distribution of Fig.\ \ref{fig4} , FP1 and FP4 cases deviate from
each other with the increase of $p_t$ which is obviously due to the
enhancement of cross sections in the FP4 case. It is interesting to
see that the difference of results between with FP3 and without
$p_{NN}$ limit can even be also detected from Fig.\ \ref{fig4}, when
$p_t \gtrsim 0.6$ GeV c$^{-1}$. A beam-energy scan of the elliptic
flow under various momentum conditions might be useful for giving
further constraints on cross sections in medium.

\begin{figure}
\includegraphics[angle=0,width=0.8\textwidth]{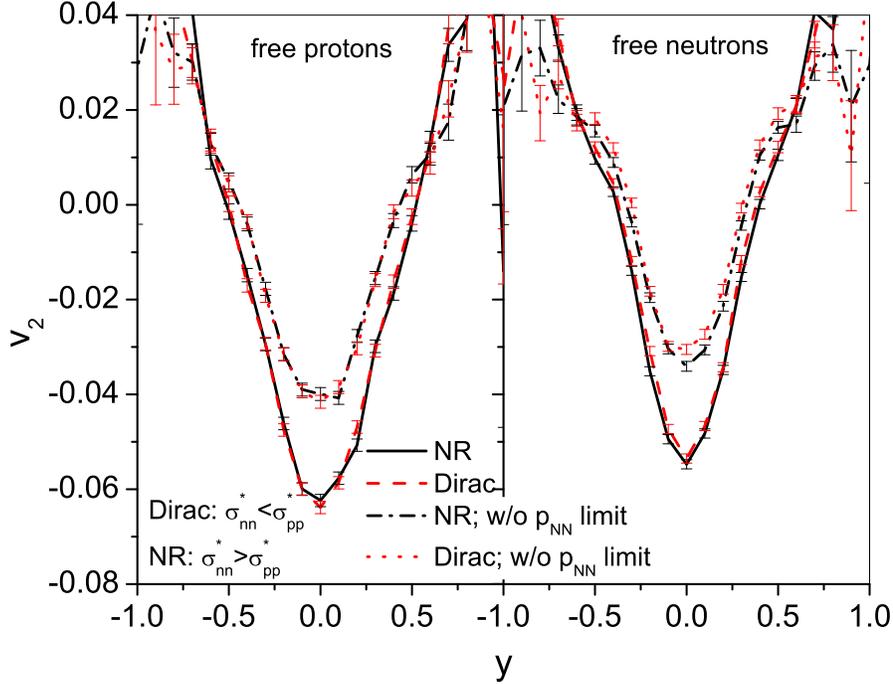}
\caption{Rapidity dependence of $v_2$ of protons (left plot) and
neutrons (right plot) with NR- and Dirac-typed splitting. For each
typed splitting effect, the flow is calculated with and without the
momentum dependence on the density dependent term $F_u$.}
\label{fig6}
\end{figure}

Now that the elliptic flow can be taken as a sensitive probe for the
momentum dependence of the medium modifications of cross sections,
it is supposed that it might also be a good candidate for detecting
the splitting effect probably shown in the NN elastic cross
sections. Fig.\ \ref{fig6} shows the rapidity dependence of $v_2$ of
protons (left plot) and neutrons (right plot) with NR- and
Dirac-type mass-splitting. In the NR case, $m_n^* > m_p^*$ so that
$\sigma^*_{nn} > \sigma^*_{pp}$, while in the Dirac case, the trend
is on the contrary \cite{split}. The detailed forms of the splitting
effect $F_\alpha$ on $\sigma^*_{nn}$ and $\sigma^*_{pp}$ had been
discussed in Ref.\ \cite{Li:2006ez}. In Fig.\ \ref{fig6} the flows
without momentum dependence of $F_u$ are also calculated for
comparison with the ones having momentum constraint. First of all,
it is seen that the $p_{NN}$ limit plays strong role on the final
elliptic flow while the splitting effect does not. For protons, the
flow with the NR-typed splitting is slightly larger than that with
the Dirac case at mid-rapidity because of a bit smaller
$\sigma^*_{pp}$ than $\sigma^*_{nn}$, while for neutrons, the
inverse observation is made which is certainly due to the same
reason. Finally, it is also interesting to find that the splitting
effect of elastic cross sections on the elliptic flow is too small
to affect the sensitivity of the elliptic flow to the mass-splitting
effect in the mean field calculations claimed in Refs.
\cite{DiToro:2006hd,DiToro:2008zm}. Actually in that work these
results were obtained just on the basis of pure mean field effects,
i.e. different momentum dependence of the symmetry potentials for
neutrons and protons, without changing the elastic cross sections.

To summarize, in order to gain deep insight into the dynamics of
particles in the nuclear medium at SIS energies, the momentum
dependence of the medium modifications on nucleon-nucleon elastic
cross sections is analyzed based on microscopic transport theories
and numerically investigated with an updated UrQMD microscopic
transport model in which the EOS and the medium modified cross
sections had been considered and examined before. The
semi-peripheral Au+Au reaction at $E_b=400A$ MeV is adopted since it
produces large negative collective flows. It is found that the
uncertainties originating from the momentum dependence on medium
modifications of cross sections, such as the slope at moderate
relative momenta as well as the possible enhancement of cross
sections at high momenta, influence the emission of free nucleons as
well as their flows. Among these, the elliptic flow is seen to be
sensitively dependent on the detailed forms of the momentum
constraint on cross sections. However, the flow is still insensitive
to the splitting effect of the neutron-neutron and proton-proton
cross sections in the isospin-asymmetrized nuclear medium.
This result can be partially related to the naive $(m^*/m)^2$ scaling
of the cross sections. The mass-splitting will be then not affecting
the dominant $(n,p)$ collisions.

In order to pin down the mass-splitting effect obtained in the mean
field calculation at high beam energies and nuclear densities by
exploring the elliptic flow of nucleons or light clusters, it is
quite necessary to dig deeper into the momentum dependence on medium
modifications of the cross sections. Moreover, the in-medium
modification of the angular distributions should also be properly
accounted for \cite{fuchs01,santini05}. A good check would be to
test the effect on the stopping, i.e. on the longitudinal and
transverse rapidity distributions.

In the next step, using the self-consistent RBUU theory, we plan also to
investigate more systematically the energy dependence of the NN
elastic cross sections in the neutron-rich nuclear medium within the
reduced density region $u \lesssim 3$ and the temperature region $T
\lesssim 100$ MeV \cite{Oeschler:2009ab}.

\section*{Acknowledgments}
This work is done partly during the workshop ``Relativistic
many-body problems for heavy and superheavy nuclei'' held in Beijing
in the summer of 2009, we would like to thank the financial support
from KITPC institute during the workshop. We thank P. Danielewicz
for useful discussions. We acknowledge support by the Frankfurt
Center for Scientific Computing (CSC). The work is supported in part
by the key project of the Ministry of Education of China under grant
No. 209053 and the National Natural Science Foundation of China
under grant No. 10675046.


\newpage
\begin{thebibliography}{99}

\bibitem{Li:2008gp}
  B.~A.~Li, L.~W.~Chen and C.~M.~Ko,
  Phys.\ Rept.\  {\bf 464}, 113 (2008).

\bibitem{Fuchs:2007vt}
  C.~Fuchs,
  J.\ Phys.\ G {\bf 35}, 014049 (2008).

\bibitem{Danielewicz:2002pu}
  P.~Danielewicz, R.~Lacey and W.~G.~Lynch,
  Science {\bf 298}, 1592 (2002).

\bibitem{Liu:2001iz}
  B.~Liu, V.~Greco, V.~Baran, M.~Colonna and M.~Di Toro,
  Phys.\ Rev.\ C {\bf 65}, 045201 (2002)



\bibitem{Mao:1994zza}
  G.~Mao, Z.~Li, Y.~Zhuo, Y.~Han and Z.~Yu,
  Phys.\ Rev.\  C {\bf 49}, 3137 (1994).

\bibitem{Mao:1997gr}
  G.~Mao, L.~Neise, H.~Stoecker, W.~Greiner and Z.~Li,
  Phys.\ Rev.\  C {\bf 57}, 1938 (1998).

\bibitem{Li:2003vd}
  Q.~Li, Z.~Li and E.~Zhao,
  Phys.\ Rev.\  C {\bf 69}, 017601 (2004).

\bibitem{Sammarruca:2005ch}
  F.~Sammarruca,
  arXiv:nucl-th/0506081.

\bibitem{Sammarruca:2005tk}
  F.~Sammarruca and P.~Krastev,
  arXiv:nucl-th/0509011.

\bibitem{Li:2005jy}
  B.~A.~Li and L.~W.~Chen,
  Phys.\ Rev.\  C {\bf 72}, 064611 (2005).


\bibitem{Larionov:2003av}
  A.~B.~Larionov and U.~Mosel,
  Nucl.\ Phys.\  A {\bf 728}, 135 (2003).

\bibitem{Prassa:2007zw}
  V.~Prassa, G.~Ferini, T.~Gaitanos, H.~H.~Wolter, G.~A.~Lalazissis and M.~Di Toro,
  Nucl.\ Phys.\  A {\bf 789}, 311 (2007).

\bibitem{Li:2000sha}
  Q.~Li, Z.~Li and G.~Mao,
  Phys.\ Rev.\  C {\bf 62}, 014606 (2000).

\bibitem{Li:2003ai}
  Q.~Li and E.~Zhao,
  Mod.\ Phys.\ Lett.\  A {\bf 18} (2003) 2713.

\bibitem{Mao2005}
G.~Mao, {\it Relativistic Microscopic Quantum Transport Equation}
(NOVA Science Publishers, New York, 2005).



\bibitem{Li:2006ez}
  Q.~Li, Z.~Li, S.~Soff, M.~Bleicher and H.~Stoecker,
  J.\ Phys.\ G {\bf 32}, 407 (2006).

\bibitem{DiToro:2006hd}
  M.~Di Toro {\it et al.},
  Nucl.\ Phys.\  A {\bf 787}, 585 (2007).

\bibitem{DiToro:2008zm}
  M.~Di Toro {\it et al.},
  Progr.\ Part.\ Nucl.\ Phys.\ {\bf 62}, 389 (2009).

\bibitem{Mao:1994et}
  G.~Mao, Z.~Li, Y.~Zhuo, Y.~Han, Z.~Yu and M.~Sano,
  Z.\ Phys.\  A {\bf 347} (1994) 173.

\bibitem{Li:1993ef}
  G.~Q.~Li and R.~Machleidt,
  Phys.\ Rev.\  C {\bf 49}, 566 (1994).

\bibitem{Giansiracusa:1996zz}
  G.~Giansiracusa, U.~Lombardo and N.~Sandulescu,
  Phys.\ Rev.\  C {\bf 53}, R1478 (1996).


\bibitem{Li:2005gf}
  Q.~Li, Z.~Li, S.~Soff, M.~Bleicher and H.~St\"ocker,
  J.\ Phys.\ G: Nucl.\ Part.\ Phys.\ {\bf 32}, 151 (2006)
\bibitem{Li:2008bk}
  Q.~Li and M.~Bleicher,
  J.\ Phys.\ G {\bf 36}, 015111 (2009).

\bibitem{Aichelin:1986wa}
  J.~Aichelin and H.~St\"ocker,
  Phys.\ Lett.\  B {\bf 176}, 14 (1986).

\bibitem{Sorge:1989dy}
  H.~Sorge, H.~St\"ocker and W.~Greiner,
  Annals Phys.\  {\bf 192}, 266 (1989).

\bibitem{Bratkovskaya:2004kv}
  E.~L.~Bratkovskaya {\it et al.},
  Phys.\ Rev.\ C {\bf 69}, 054907 (2004).

\bibitem{Zhang:2007gd}
  Y.~Zhang, Z.~Li and P.~Danielewicz,
  Phys.\ Rev.\  C {\bf 75}, 034615 (2007).


\bibitem{Kru85}H. Kruse, B.V. Jacak, J.J. Molitoris, G.D. Westfall, H. St\"ocker, Phys. Rev. C {\bf 31}, 1770
(1985).

\bibitem{split}
We note that in a full relativistic approach a connection can be
worked out between the Dirac and NR (Non-Relativistic) evaluation.
The relation is however strongly affected by the poorly known
momentum dependence of the nucleon self-energies, see Sect.6.3.1 of
Ref.\ \cite{baranPR} and the detailed Dirac-Brueckner calculation of
Ref.\ \cite{vandalen05}. In this paper we have followed the rough
choices described in the text in order to test possible effects on
reaction observables.

\bibitem{baranPR}
V.~Baran, M.~Colonna, V.~Greco, M.~Di Toro, Phys.\ Rep.\ {\bf 410},
335 (2005).

\bibitem{vandalen05}
E.~N.~E.~van Dalen, C.~Fuchs, A.~Faessler, Phys.\ Rev.\  C {\bf 72},
065803 (2005).

\bibitem{fuchs01}
C.~Fuchs, A.~Faessler, M.~El-Shabshiry, Phys. Rev. C {\bf 64},
024003 (2001).

\bibitem{santini05}
E.~Santini, T.~Gaitanos, M.~Colonna, M.~Di Toro, Nucl. Phys. A {\bf
756}, 468 (2005).

\bibitem{Oeschler:2009ab}H.~Oeschler, H.~G.~Ritter and N.~Xu,
 arXiv:nucl-ex/0908.1771.

\end{thebibliography}
\end{document}